\documentclass{ab}

\usepackage{graphicx}
\usepackage{natbib}
\usepackage{lscape}

\def\degr{\hbox{$^\circ$}}
\def\arcmin{\hbox{$^\prime$}}
\def\arcsec{\hbox{$^{\prime\prime}$}}

\begin{document}

\title{Spectroscopy of H~II Regions in the Late-Type Spiral Galaxy NGC~6946}

\author{A.~S.~Gusev,$^1$
        F.~H.~Sakhibov,$^2$
        S.~N.~Dodonov$^3$}

\institute{$^1$ Sternberg State Astronomical Institute, Lomonosov Moscow State
University, Moscow, 119992 Russia \\
           $^2$ University of Applied Sciences of Mittelhessen, Friedberg, 61169
Germany \\
           $^3$ Special Astrophysical Observatory, Russian Academy of Sciences, 
Nizhnii Arkhyz, 369167 Russia}

\date{Received June 1, 2012; in final form, October 2, 2012}
\offprints{A.~S.~Gusev, \email{gusev@sai.msu.ru}}

\titlerunning{Spectroscopy of H~II Regions in the Late-Type Spiral NGC~6946}
\authorrunning{Gusev et al.}

\abstract{We present the results of spectroscopy of 39 H~II regions in the 
spiral galaxy NGC~6946. The spectral observations were carried out at the 
6-m BTA telescope of the SAO RAS with the SCORPIO focal reducer in the 
multi-slit mode with the dispersion of 2.1\AA/px and spectral resolution of 
10\AA. The absorption estimates for 39 H~II regions were obtained. Using the 
''strong line'' method (NS-calibration) we determined the electron 
temperature, and the abundances of oxygen and nitrogen for 30 H~II regions. 
The radial gradients of O/H and N/H were constructed. \\

{\bf DOI:} 10.1134/S1990341313010045 \\

}

\maketitle

\section{INTRODUCTION} 

This work is part of our research aimed at determining the physical 
parameters of H~II regions, and studying the processes of star formation 
in spiral and irregular galaxies based on the spectroscopic and photometric 
data. In \citet{gusev2012} we have analyzed the results of spectroscopic 
observations of the H~II regions in six spiral galaxies. The NGC~6946 galaxy, 
for which the largest sample of objects was obtained has a number of 
features requiring a separate discussion. We have therefore dedicated a 
separate paper to the analysis of spectral observations of its H~II regions. 

A nearby late-type spiral galaxy NGC~6946, turned almost face-on to the 
observer from Earth has been actively studied for more than half a century. 
Numerous H~II regions and a large number of registered supernovae make it a 
suitable object for the study of star formation processes in the modern 
epoch. Given all that, the reduction and interpretation of spectroscopic data 
poses a serious problem \citep{efremov2011} since the NGC~6946 is located at 
a low galactic latitude \citep[$b = 11.7\degr$,][]{paturel2003}. Accounting 
for the effect of the Milky Way requires particular care during the data 
reduction. 

The spectral studies of H~II regions in the galaxies provide information on 
the chemical composition and its variation with galactocentric distance in 
the galactic disks. Despite the fact that there is a large number of spectral 
observations of H~II regions in other galaxies \citep{zaritsky1994,roy1996,
vanzee1998,dutil1999,kennicutt2003,bresolin2005,bresolin2009}, in NGC~6946 
they were carried out only for nine H~II regions 
\citep{mccall1985,ferguson1998}. A number of investigations were devoted to 
the detailed studies of certain large H~II complexes, located in the 
peripheral parts of the galactic disk; four complexes, located in a close 
proximity in the eastern part of the galaxy were studied using the panoramic 
spectroscopy in \citet{garcia2010}; the spectroscopic studies of the stellar 
complex located westerly were conducted in \citet{efremov2002,efremov2007}. 
In 1992, \citet{belley1992} published the estimates of the chemical 
composition for 166 H~II regions in NGC~6946 based on the spectrophotometry 
of four emission lines (H$\alpha$, H$\beta$, [N\,{\sc ii}] and 
[O\,{\sc iii}]), conducted with the narrowband interference filters. However, 
the ''strong line'' method used by the authors to calibrate the oxygen 
abundance does not give an unambiguous estimate of the true oxygen and 
nitrogen abundance \citep[see][]{pilyugin2003,ellison2008,lopezsanchez2010}. 
Since all the lines were calibrated separately, the systematic 
underestimation or overestimation of fluxes in different lines, noted in 
\citet{belley1992} leads to the systematic errors in the relative 
intensities, as seen on the diagnostic BPT diagram \citep{baldwin1981}, 
given in \citet{belley1992}, where a significant number of H~II regions have 
revealed a nonthermal nature of the emission lines. Therefore, the relative 
intensities of the lines of hydrogen, oxygen, nitrogen and sulfur, measured 
in the present study from the analysis of an integral spectrogram, covering 
the range from 3200 to 7000\AA \, for 30 objects in NGC~6946 is an essential 
complement to the sample of nine H~II regions, earlier obtained by two teams 
of researchers \citep{mccall1985,ferguson1998}.

An important feature of the galaxy was noted by \citet{boomsma2008}, who 
observed it in the 21-cm line. The authors have detected large 
irregularities in the spatial distribution of H~I and in the velocity field. 
One hundred and twenty-one cavities of neutral hydrogen were identified; 
their sizes reach up to 2.5 kpc. More than $4\%$ of H~I by mass has a velocity 
which differs by over 50 km/s from the circular velocity at a given distance 
from the center. Large deviations from circular velocities are due to the 
presence of the first mode (along with the second mode) of the spiral 
density wave, which occur in the Fourier analysis of the two-dimensional 
velocity field \citep{sakhibov2004}. Therefore, the NGC~6946 is an isolated 
spiral galaxy with a classical structure, having, however, a peculiar 
distribution of neutral gas, and is hence a very interesting object to study 
the physical parameters of the H~II regions and the features of star 
formation process.

\begin{table}
\begin{center}
\caption{Main characteristics of NGC~6946.}
\label{table:data}
\begin{tabular}{|l|c|}
\hline\hline
Parameter         & Value \\
\hline
Type              & SABc \\
$\alpha_{2000}$   & $20^h34^m52.75^s$ \\
$\delta_{2000}$   & $+60\degr09\arcmin13.6\arcsec$ \\
$m_B$             & $9.75^m$ \\
$M_B^{0, i}$      & $-20.68^m$ \\
$i$               & $31\degr$ \\
P.A.              & $62\degr$ \\
$V$               & 46 km/s \\
$R^c_{25}$        & $7.74\arcmin$ \\
$R^c_{25}$        & 13.28 kpc \\
$d$               & 5.9 Mpc \\
$A(B)_{Gal}$      & $1.48^m$ \\
$A(B)_i$          & $0.04^m$ \\[1mm]
\hline
\end{tabular}
\end{center}
\end{table}

The main characteristics of the galaxy: its type, $\alpha_{2000}$ and 
$\delta_{2000}$ coordinates, apparent magnitude $m_B$, absolute magnitude 
$M_B^{0, i}$, corrected for the galactic extinction and absorption caused by 
the inclination of NGC~6946, inclination $i$, position angle P.A., radial 
velocity $V$, the radius from the isophotes $25^m$ in the $B$-band 
$R^c_{25}$, the distance $d$, galactic extinction $A(B)_{Gal}$ and absorption 
$A(B)_i$, caused by the inclination of NGC~6946 are presented in 
Table~\ref{table:data}. The distance $d$ in the table is listed according to 
\citet{karachentsev2000}, the other parameters were adopted from the 
HyperLeda database \citep{paturel2003}. Note that unlike most of researchers, 
we use the radius $R^c_{25}$ by the isophote $25^m$, corrected for the 
galactic extinction and absorption caused by the inclination of NGC~6946.

Given the large extinction of our Galaxy in the direction of NGC~6946, the 
corrected $R^c_{25}$ value is by $40\%$ greater than $R_{25}$, not corrected 
for extinction. Hence, all the galactocentric distances in the scale of 
$R^c_{25}$, obtained in this study may differ from the data of other 
authors, who used the value of $R_{25}$. Further in the text, we omit the 
$c$ index, implying everywhere the $R_{25}$ value corrected for the 
absorption (see Table~\ref{table:data}).

In this paper, we did not give consideration to the H~II regions with the 
absorption lines in the spectrum. Such regions were excluded from our 
further research (such as the famous 
\citep{hodge1967,elmegreen2000,efremov2002,efremov2007} giant peculiar star 
complex located westwards in the galaxy).

\section{OBSERVATIONS AND DATA REDUCTION}

\subsection{Observations}

\begin{figure}
\vspace{1.0cm}
\centerline{\includegraphics[width=8.0cm]{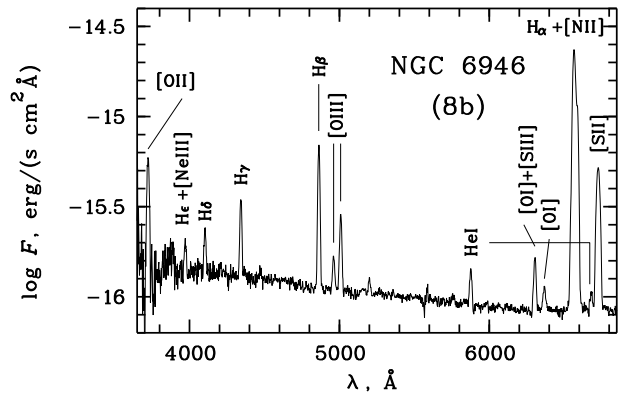}}
\caption{The ''b'' spectrum of the eighth H~II region of the galaxy.
\label{figure:spect}}
\end{figure}

Spectral observations were conducted in 2007 at the 6-m BTA telescope of 
the Special Astrophysical Observatory of the Russian Academy of Sciences 
(SAO RAS) with the SCORPIO focal reducer \citep[for a detailed description 
of the device, see][]{afanasiev2005} in the multi-slit mode. The detector 
used was a CCD camera EEV 42-40. The size of the chip amounts to 
$2048 \times 2048$ px, which provides a field of view of $6\arcmin$ given 
the image scale of $0.178\arcsec$ per pixel. In the multi-slit mode the 
SCORPIO instrument has 16 movable slits, installed in the focal plane and 
relocating in the field of $2.9\arcmin \times 5.9\arcmin$. The size of the 
slits is $1.5\arcsec \times 18\arcsec$, the distance between the centers of 
adjacent slits amounts to $22\arcsec$. The log of spectral observations at 
the BTA is given in Table~\ref{table:obsspec}.

During the observations we used the VPHG550G grism with the dispersion of 
2.1\AA/px and spectral resolution of 10\AA. This grism allows to register 
the radiation in the range of 3100–-7300\AA, where the range edges vary 
depending on the position of the slit. The spectral range of the grism 
allows to obtain the lines from 
[O\,{\sc ii}]$\lambda$3727+$\lambda$3729\AA \, to 
[S\,{\sc ii}]$\lambda$6717+$\lambda$6731\AA \, in one spectrum 
(Fig.~\ref{figure:spect}).

The choice of the H~II regions for spectral observations was made based on 
the images of the galaxy in the $B$ and H$\alpha$ filters, earlier obtained 
at the 1.5-m telescope of the Maidanak Observatory (Uzbekistan) with the 
angular resolution of about $1\arcsec$ (unpublished data). We have selected 
the bright (both in the $B$ and H$\alpha$ bands) regions with the angular 
sizes from 2 to $5\arcsec$, located in a wide range of galactocentric 
distances. For each of the three sets of slit positions, from six to eight 
15-minute exposures were obtained (Table~\ref{table:obsspec}). After each 
exposure the slit positions were shifted rightwards-–leftwards along the slit 
in the increments of 20 px. This made it possible to obtain the spectra of 
several nearby H~II regions in a single slit. However, due to the shifts the 
total time of exposure for a number of individual H~II regions was less than 
the total exposure time, specified in Table~\ref{table:obsspec}.

For the standard reduction and data calibration, at the beginning and end 
of each set of observations of NGC~6946, the images with the zero exposure 
(bias), flat field, the spectra of the helium-neon-argon lamp and the 
comparison star were obtained.

\begin{table}
\begin{center}
\caption{Log of spectral observations.}
\label{table:obsspec}
\begin{tabular}{|c|c|c|c|c|}
\hline\hline
Date & Set of    & Exposure, & Seeing, & Air \\
     & slit      & s         & arcsec  & mass \\
     & positions &           &         &  \\
\hline
Sep 06, 2007 & 1 & $900\times6$ & 1.2 & 1.08 \\
             & 2 & $900\times6$ & 1.2 & 1.57 \\
Sep 07, 2007 & 3 & $900\times8$ & 1.6 & 1.19 \\[1mm]
\hline
\end{tabular}
\end{center}
\end{table}

\subsection{Data Reduction}

Further data reduction was carried out at the SAI MSU via a standard 
procedure using the ESO MIDAS image processing system. The main processing 
steps included: removal of traces of cosmic ray particles, identifying and 
correcting the data for the bias and flat field; conversion to the 
wavelength scale using the spectrum of a He-Ne-Ar lamp; normalization of 
fluxes by the intensity of the central (eighth) slit; background subtraction; 
conversion of the instrumental fluxes into absolute, using the observational 
data of spectrophotometric standard stars and the correction for the 
atmospheric extinction; the integration of two-dimensional spectra in the 
selected apertures for obtaining the one-dimensional spectra of the 
individual H~II region; addition of spectra for each region. An example of 
the resulting spectrum is shown in Fig~\ref{figure:spect}.

In the transition to one-dimensional spectra we integrated the 
two-dimensional spectra in the apertures, corresponding to the areas, where 
the bright emission lines of the H~II regions were recognizable above the 
noise. The aperture size approximately corresponds to the diameter of an 
individual H~II region along the slit position angle.

The continuum was predetermined and subtracted from the spectra to measure 
the fluxes in the emission lines. The star BD+25$\degr$4655 \citep{oke1990} 
was used as a spectrophotometric standard. To calculate the atmospheric 
extinction coefficient and correct for the atmospheric extinction, the 
results of astroclimatic measurements were used along with the observations 
of BD+25$\degr$4655 \citep{kartasheva1978}. To separate the fluxes of the 
blended emission lines, they (the doublets or triplets) were simultaneously 
described by two or three Gaussians.

\begin{figure*}
\centerline{\includegraphics[width=16cm]{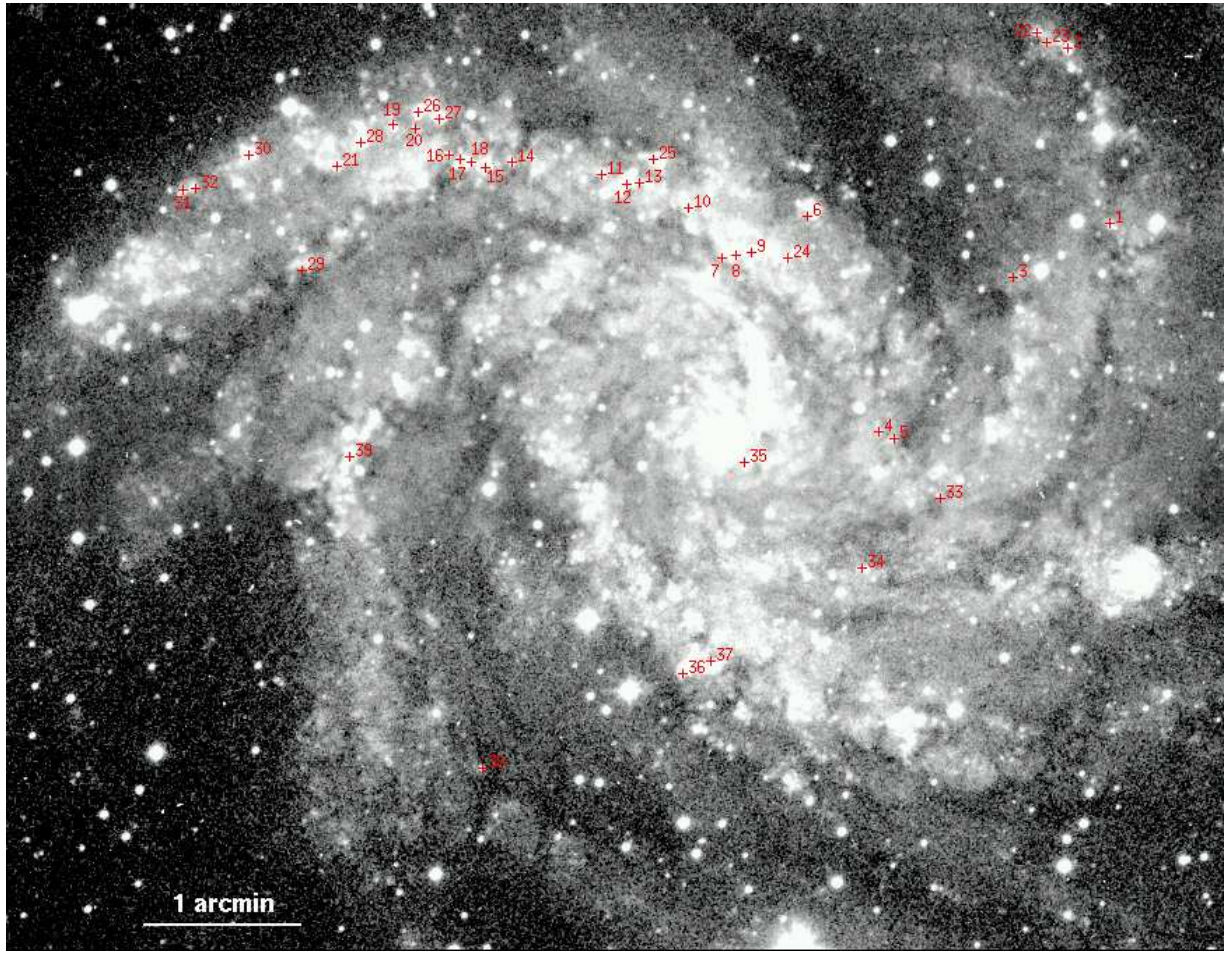}}
\caption{The image of the galaxy in the $B$-band. The positions of H~II 
regions are marked. The numbers correspond to the index number of the region 
in Table~\ref{table:poshii}. North is on top, east is to the left.
\label{figure:map}}
\end{figure*}

All in all, we have obtained the spectra of 39 H~II regions 
(Fig.~\ref{figure:map}). The regions observed twice are marked in 
Tables~\ref{table:poshii}, \ref{table:flux1}, \ref{table:abun} by letters 
''a'' and ''b'' for the first and second obtained spectra, respectively.

The line intensity measurement errors include several components. The first 
source of errors is associated with the Poisson photon statistics in the 
line flux. The second component is caused by the error in determining the 
continuum level under the emission line and provides the main contribution 
to the total error for single lines. The third source of errors is associated 
with the accuracy of determining the spectral sensitivity curve, it is 
significant (over $1\%$) in the short-wavelength spectral region for the 
wavelengths of $\lambda < 4000$\AA. The last source of error, which is 
important for blended lines is due to the blend approximation errors by the 
Gaussians. The total error of line intensity was determined by quadratic 
summation of all the error components. These full errors were then correctly 
converted into the errors of calculated parameters by standard formulae.

Note that the absolute values of fluxes in the emission lines obtained for 
one and the same H~II region may significantly differ from one set to another. 
This is caused by the seeing variation and the error of slit pointing on the 
object. Also note that the angular sizes of the observed H~II regions are by 
2–-3 times larger than the slit width, and the regions themselves can 
possess a complex internal structure. For these reasons the absolute values 
of fluxes, obtained for the H~II regions observed twice may be very different 
from each other (see Table~\ref{table:poshii}), while the line flux ratios 
for these regions are almost everywhere identical within the error margins 
(Table~\ref{table:flux1}).

We estimated the equivalent widths (EW) of the H$\alpha$ and H$\beta$ 
emission lines from the spectra of H~II regions, taking into account the 
continuum (Fig.~\ref{figure:spect}). Such spectra were constructed by 
subtracting from the spectrum of the H~II region the spectrum of the 
surrounding galactic disk background. This allows us to eliminate the 
contribution of stars and gas of the NGC~6946 disk in the radiation, coming 
from the H~II region, and eliminates the problem of accounting for the 
contribution of the diffuse extraplanar ionized gas of the Galaxy, 
considered in detail by \citet{efremov2011}. For some H~II regions the 
resulting level of the continuum appeared to by very small (close to zero), 
causing the obtained EW values to be unrealistically high, and the error 
estimates to be huge. Such data are not included in 
Table~\ref{table:poshii}.

Table~\ref{table:poshii} lists the coordinates, deprojected galactocentric 
distances $r$, equivalent H$\alpha$ and H$\beta$ line widths and 
$F$(H$\beta$) fluxes of the H~II regions, not corrected for absorption.

Table~\ref{table:flux1} lists the normalized for H$\beta$ and corrected for 
the interstellar extinction of light relative intensities of the 
[O\,{\sc ii}]$\lambda$3727+$\lambda$3729\AA, 
[O\,{\sc iii}]$\lambda$5007\AA, [N\,{\sc ii}]$\lambda$6584\AA, 
and [S\,{\sc ii}]$\lambda$6717+$\lambda$6731\AA \, lines. The account of the 
extinction of the gas radiation emission lines was carried out by the 
observed values of the Balmer decrement in the spectra of the investigated 
objects. We used the theoretical ratio of the H$\alpha$/H$\beta$ lines from 
\citet{osterbrock1989} for the case B: recombination at the electron 
temperature of $10^4$~K, and analytical approximation of \citet{izotov1994} 
of the Whitford interstellar reddening law. We took the equivalent width of 
the hydrogen absorption lines EW$_a$($\lambda$) to be equal to 2\AA \, for all 
objects. According to \citet{mccall1985}, this value is the average for the 
H~II regions. For the lines of other chemical elements EW$_a$($\lambda$) = 0.

Determining the errors of the corrected for interstellar extinction of 
light relative intensities of lines, presented in Table~\ref{table:flux1}, 
we took into account the errors of intensity of the corresponding line, the 
H$\beta$ line, and the coefficient of absorption $c$(H$\beta$).

\section{ANALYSIS}

\subsection{Abundance of Chemical Elements}

\begin{figure}
\vspace{6.0mm}
\centerline{\includegraphics[width=7.0cm]{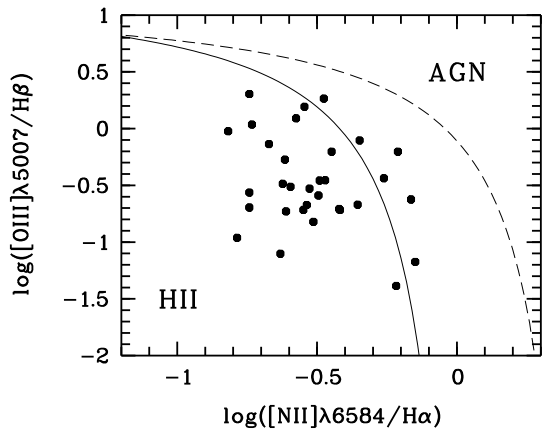}}
\caption{The BPT diagram for the H~II regions, studied in the present paper 
(black circles). The solid line separating the classical H~II regions from 
the objects with nonthermal emission spectra (AGN) is calculated according 
to \citet{kewley2001}. The dashed line is the curve, adopted from 
\citet{kauffmann2003}.
\label{figure:bpt}}
\end{figure}

Take a look at the positions of our objects in the diagnostic BPT diagram 
(Fig.~\ref{figure:bpt}). The figure clearly shows that all the objects are 
located in the region, where the emission lines are emerging as a result of 
thermal radiation, implying that all the objects are the classical H~II 
regions. Based on this, all these objects are taken for the further analysis 
of chemical elements.

\begin{table*}
\begin{center}
\caption{Parameters of the H~II regions, equivalent widths of the H$\alpha$, 
H$\beta$ lines, and the H$\beta$ line fluxes.}
\label{table:poshii}
\begin{tabular}{|c|c|c|c|c|c|c|c|}
\hline\hline
H~II    & Set & Number from         & Coordinates, & $r$, & EW(H$\beta$), & 
EW(H$\alpha$), & $F$(H$\beta$), \\
region &     & \citet{belley1992}, & arcsec       & kpc & \AA           & 
\AA            & 10$^{-16}$ \\
       &     & {\bf \citet{mccall1985}}, &        &     &               & 
               & erg/(s$\cdot$cm$^{2}$) \\
       &     & {\it \cite{ferguson1998}} &       &     &               &
               &            \\
\hline
 1  & 1 & 116 & 79.7N, 146.5W & 5.34 & --- & --- & $6.63\pm0.38$ \\
 2  & 1 & 143, {\bf 5} & 146.9N, 130.5W & 6.51 & $81.4\pm3.4$ & $519.0\pm24.7$ & $84.06\pm1.33$ \\
 3  & 1 & --- & 58.9N, 110.0W & 3.99 & $14.9\pm5.4$ & $106.5\pm30.1$ & $1.23\pm0.20$ \\
 4  & 1 & --- &  0.3S, 58.8W & 1.75 & $27.0\pm3.5$ & $105.6\pm8.5$ & $5.97\pm0.23$ \\
5a  & 1 & --- &  2.7S, 64.6W & 1.91 & $15.8\pm5.0$ & $80.5\pm13.9$ & $1.85\pm0.24$ \\
5b  & 2 & --- &  2.7S, 64.6W & 1.91 & $9.6\pm1.7$ & $61.2\pm6.8$ & $3.61\pm0.33$ \\
 6  & 1 & 149 & 82.6N, 31.6W & 2.95 & $9.5\pm0.4$ & $92.2\pm3.8$ & $21.03\pm0.60$ \\
 7  & 1 & --- & 66.3N, 1.2E & 2.14 & $1.1\pm0.4$ & $37.2\pm8.8$ & $0.30\pm0.07$ \\
8a  & 1 & {\bf 1} & 67.7N, 4.1W & 2.21 & $27.9\pm2.1$ & $151.1\pm11.4$ & $8.86\pm0.23$ \\
8b  & 2 & {\bf 1} & 67.7N, 4.1W & 2.21 & $84.0\pm4.3$ & $645.8\pm38.3$ & $92.95\pm1.52$ \\
 9  & 1 & 150 & 68.5N, 10.2W & 2.28 & $102.1\pm6.0$ & $939.3\pm77.3$ & $33.02\pm0.54$ \\
10  & 1 &  10 & 85.5N, 14.0E & 2.75 & $15.0\pm0.8$ & $117.7\pm5.5$ & $22.97\pm0.65$ \\
11  & 1 & --- & 98.6N, 46.8E & 3.32 & $61.8\pm8.6$ & $577.0\pm80.3$ & $29.52\pm0.84$ \\
12  & 1 &   8 & 94.9N, 37.5E & 3.13 & $18.9\pm0.8$ & $173.9\pm9.0$ & $20.31\pm0.43$ \\
13  & 1 &   6 & 95.4N, 32.4E & 3.12 & $49.1\pm5.7$ & $264.3\pm28.2$ & $13.50\pm0.40$ \\
14  & 1 &  24 & 103.4N, 81.0E & 3.87 & $158.5\pm25.9$ & $997.9\pm160.3$ & $24.82\pm0.45$ \\
15a & 1 &  25 & 101.0N, 91.4E & 3.98 & $103.9\pm30.3$ & $771.1\pm273.9$ & $13.75\pm0.38$ \\
15b & 2 &  25 & 101.0N, 91.4E & 3.98 & $85.0\pm19.7$ & $525.6\pm93.9$ & $21.21\pm0.67$ \\
16  & 1 &  29 & 106.1N, 105.0E & 4.34 & $136.0\pm53.2$ & $664.4\pm134.7$ & $19.44\pm0.58$ \\
17  & 1 &  28 & 104.2N, 101.0E & 4.22 & $50.2\pm4.6$ & $385.0\pm26.7$ & $17.14\pm0.37$ \\
18  & 2 &  26 & 103.1N, 96.7E & 4.12 & $44.3\pm4.2$ & $179.7\pm11.9$  & $27.25\pm0.65$ \\
19  & 1 &  16 & 117.8N, 126.3E & 5.00 & $113.2\pm13.7$ & $1055.1\pm107.0$ & $23.54\pm0.43$ \\
20  & 1 & --- & 116.2N, 118.0E & 4.81 & --- & --- & $10.67\pm0.36$ \\
21  & 1 &  19 & 101.8N, 147.6E & 5.14 & $107.7\pm21.7$ & $570.4\pm55.1$ & $48.75\pm1.11$ \\
22  & 2 & 146 & 153.0N, 118.8W & 6.44 & $113.9\pm22.8$ & $304.3\pm14.4$ & $17.37\pm0.46$ \\
23  & 2 & 145 & 149.3N, 122.5W & 6.41 & $73.6\pm9.0$ & $333.7\pm28.6$ & $16.56\pm0.38$ \\
24  & 2 & 151 & 66.3N, 23.8W & 2.35 & $8.5\pm3.0$ & $30.4\pm3.5$ & $4.00\pm0.87$ \\[1mm]
25  & 2 &   5 & 104.5N, 27.1E & 3.38 & $39.7\pm2.7$ & $416.9\pm35.1$ & $39.02\pm0.88$ \\
26  & 2 &  15 & 122.3N, 116.7E & 4.92 & $27.5\pm1.2$ & $158.5\pm6.5$ & $26.38\pm0.49$ \\
27  & 2 &  14 & 119.9N, 108.7E & 4.72 & $92.8\pm8.7$ & $691.4\pm47.4$ & $59.51\pm1.03$ \\
28  & 2 &  17 & 110.6N, 138.6E & 5.10 & $7.1\pm1.2$ & $113.0\pm15.4$ & $4.05\pm0.34$ \\
29  & 2 & --- & 61.8N, 161.2E & 4.95 & $13.3\pm2.1$ & $114.8\pm12.5$ & $9.55\pm0.65$ \\
30  & 2 &  39, {\bf 4}, {\it A} & 105.8N, 181.2E & 6.00 & $83.9\pm3.4$ & $560.4\pm15.1$ & $144.84\pm2.28$ \\ 
31  & 2 & --- & 92.5N, 206.3E & 6.47 & $31.4\pm4.2$ & $315.8\pm56.9$ & $9.68\pm0.35$ \\
32  & 2 & --- & 93.3N, 201.5E & 6.35 & $13.8\pm1.1$ & $131.5\pm7.5$ & $8.55\pm0.31$ \\
33  & 3 & --- & 25.7S, 82.0W & 2.47 & $59.8\pm29.6$ & $292.1\pm95.9$ & $8.90\pm0.86$ \\
34  & 3 & --- & 52.6S, 52.4W & 2.16 & $23.0\pm1.3$ & $208.5\pm15.3$ & $33.43\pm0.78$ \\
35  & 3 & 163 & 12.1S, 7.6W & 0.43 & $6.1\pm0.7$ & $38.4\pm2.3$ & $6.39\pm0.46$ \\
36  & 3 &  78 & 92.9S, 15.9E & 3.10 & $14.7\pm0.8$ & $117.2\pm5.7$ & $42.23\pm1.12$ \\
37  & 3 & --- & 88.1S, 5.5E & 2.88 & $6.9\pm1.9$ & $83.7\pm13.0$ & $2.83\pm0.36$ \\
38  & 3 & --- & 129.1S, 92.4E & 5.29 & --- & --- & $4.63\pm0.16$ \\
39  & 3 &  49, {\bf 3} &  9.7S, 142.8E & 4.30 & $56.2\pm3.3$ & $399.9\pm21.1$ & $81.34\pm1.46$ \\[1mm]
\hline
\end{tabular}
\end{center}
\end{table*}

A number of empirical correlations between the relative intensities of 
emission lines and the chemical composition and temperature of the emitting 
gas have been proposed at different times by many researchers 
\citep[see the surveys][]{ellison2008,lopezsanchez2010}. Let us outline among 
them two recent calibrations used to determine the metallicity in the H~II 
regions. Both the ON- \citep{pilyugin2010} and NS-calibrations 
\citep{pilyuginmattsson2011} interpret the relative intensities
of strong emission lines (O$^{2+}$, O$^+$, and N$^+$ in the case of 
ON-calibration or O$^{2+}$, N$^+$, and S$^+$ in the case of NS-calibration) 
in terms of the relative oxygen and nitrogen abundances and electron 
temperatures. Since the relative intensities of 
[O\,{\sc ii}]$\lambda$3727+$\lambda$3729/H$\beta$ were measured with 
large errors, or have not been measured at all in many objects, we chose 
the NS-calibration to determine the metallicity and electron temperature of 
the gas in the studied H~II regions.

As the [O\,{\sc iii}]$\lambda$4959\AA \, line intensities were measured 
by us only for a half of H~II regions, and the 
[N\,{\sc ii}]$\lambda$6548\AA \, line intensities were determined with 
very high errors, in the calculation of the oxygen and nitrogen abundances, 
and electron temperature, we used the following ratios: 
[O\,{\sc iii}]$\lambda$4959+$\lambda$5007\AA=1.33[O\,{\sc iii}]$\lambda$5007\AA \, 
and 
[N\,{\sc ii}]$\lambda$6548+$\lambda$6584\AA=1.33[N\,{\sc ii}]$\lambda$6584\AA \,
\citep[following][]{storey2000}.

The NS-method is applicable to the regions of low density. Unfortunately, 
the sulfur [S\,{\sc ii}]$\lambda$6717\AA \, and 
[S\,{\sc ii}]$\lambda$6731\AA \, doublet lines are blended in our spectra 
and can only be isolated with a poor accuracy. Hence, we could not 
confidently determine the density of the studied objects. Nevertheless, for 
all the studied H~II regions, 
[S\,{\sc ii}]$\lambda$6717\AA/[S\,{\sc ii}]$\lambda$6731\AA$\ge$1 
(Table~\ref{table:flux1}), what corresponds to the density values of 
$N_e \le 300$~cm$^{-3}$. Thus, the estimates show that all our objects 
have low densities, which is typical of giant H~II regions, observed in 
other galaxies \citep{kennicutt1984,zaritsky1994,bresolin2005,gutierrez2010}.

\begin{figure}
\vspace{6.0mm}
\centerline{\includegraphics[width=7.0cm]{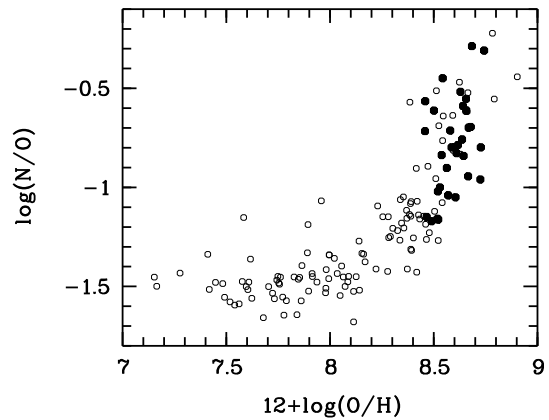}}
\caption{The O/H–-N/O diagram. The open circles represent a sample of 
well-studied H~II regions in the nearby galaxies from \citet{pilyugin2010}, 
black circles are the objects studied in this paper.
\label{figure:ohno}}
\end{figure}

The estimates of metallicity and electron temperature, obtained using the 
NS-calibration in our sample are given in Table~\ref{table:abun}. For the 
NS-calibration we used the spectra of H~II regions with reliable estimates 
of electron temperature \citep{pilyuginmattsson2011}. Therefore, the 
metallicity estimates are in a good agreement with the values given by the 
classical $T_e$-method (see Fig.~\ref{figure:ohno}). It is clear from the 
O/H–--N/O diagram (Fig.~\ref{figure:ohno}) that the objects we have studied 
in the NGC~6946 are within the lane, occupied by the most reliably 
investigated H~II regions in other galaxies. This agreement indicates the 
validity of metallicity estimates obtained for the studied objects.

\begin{table*}
\begin{center}
\caption{The absorption coefficients $c$(H$\beta$), emission line fluxes 
corrected for reddening (in the units of $I$(H$\beta$)) and the ratio of 
sulfur lines [S\,{\sc ii}]6717\AA/[S\,{\sc ii}]6731\AA.}
\label{table:flux1}
\begin{tabular}{|c|c|c|c|c|c|c|}
\hline\hline
H~II    & $c$(H$\beta$) & [O\,{\sc ii}] & [O\,{\sc iii}] & [N\,{\sc ii}] 
&[S\,{\sc ii}] & [S\,{\sc ii}]6717\AA/ \\
region  &               & 3727+3729\AA  & 5007\AA        & 6584\AA 
& 6717+6731\AA & [S\,{\sc ii}]6731\AA  \\
\hline
 1 & 1.26$\pm$0.12 & --- & 0.11$\pm$0.03 & 0.47$\pm$0.12
& 0.54$\pm$0.11 & 1.49$\pm$0.44 \\
 2 & 1.80$\pm$0.04 & --- & 0.79$\pm$0.03 & 1.29$\pm$0.11
& 0.76$\pm$0.05 & 1.20$\pm$0.18 \\
 3 & 0.54$\pm$0.34 & --- & --- & 1.29$\pm$0.67
& 1.02$\pm$0.51 & 2.54$\pm$1.31 \\
 4 & 0.19$\pm$0.09 & --- & --- & 1.04$\pm$0.17
& 0.47$\pm$0.09 & 4.80$\pm$3.69 \\
 5a & 0.83$\pm$0.28 & --- & --- & 0.96$\pm$0.44
& 0.45$\pm$0.21 & --- \\
 5b & 0.68$\pm$0.21 & --- & --- & 0.74$\pm$0.28
& --- & --- \\
 6 & 0.90$\pm$0.07 & --- & 0.35$\pm$0.04 & 0.97$\pm$0.13
& 0.56$\pm$0.07 & 1.55$\pm$0.33 \\
 7 & 1.57$\pm$0.72 & --- & --- & 3.31$\pm$3.10
& --- & --- \\
 8a & 0.60$\pm$0.07 & --- & 0.21$\pm$0.03 & 0.83$\pm$0.12
& 0.65$\pm$0.07 & 1.20$\pm$0.26 \\
 8b & 0.95$\pm$0.05 & 1.68$\pm$0.21 & 0.29$\pm$0.01 & 0.85$\pm$0.10
& 0.65$\pm$0.04 & 1.02$\pm$0.17 \\
 9 & 1.11$\pm$0.05 & 1.23$\pm$0.20 & 0.21$\pm$0.01 & 0.52$\pm$0.08
& 0.51$\pm$0.03 & 1.09$\pm$0.21 \\
10 & 0.96$\pm$0.07 & --- & 0.04$\pm$0.02 & 1.73$\pm$0.20
& 0.62$\pm$0.07 & 1.56$\pm$0.31 \\
11 & 0.92$\pm$0.07 & --- & 0.07$\pm$0.02 & 2.03$\pm$0.21
& 0.67$\pm$0.07 & 0.95$\pm$0.16 \\
12 & 1.12$\pm$0.06 & --- & --- & 0.80$\pm$0.10
& 0.55$\pm$0.05 & 1.87$\pm$0.49 \\
13 & 1.38$\pm$0.07 & --- & 0.24$\pm$0.03 & 1.97$\pm$0.22
& 0.82$\pm$0.10 & 1.38$\pm$0.31 \\
14 & 1.57$\pm$0.05 & --- & 0.19$\pm$0.02 & 0.81$\pm$0.09
& 0.46$\pm$0.04 & 1.62$\pm$0.32 \\
15a & 1.06$\pm$0.07 & --- & 0.35$\pm$0.04 & 0.92$\pm$0.12
& 0.83$\pm$0.09 & 1.18$\pm$0.21 \\
15b & 1.01$\pm$0.07 & --- & 0.33$\pm$0.04 & 0.68$\pm$0.10
& 0.29$\pm$0.04 & 1.18$\pm$0.28 \\
16 & 1.76$\pm$0.07 & --- & 0.20$\pm$0.03 & 1.09$\pm$0.13
& 0.51$\pm$0.06 & 1.34$\pm$0.24 \\
17 & 1.81$\pm$0.06 & --- & 0.22$\pm$0.02 & 1.26$\pm$0.12
& 0.50$\pm$0.04 & 1.12$\pm$0.17 \\
18 & 0.63$\pm$0.06 & --- & 0.27$\pm$0.03 & 0.52$\pm$0.08
& 0.71$\pm$0.06 & 1.64$\pm$0.30 \\
19 & 1.40$\pm$0.05 & --- & 0.53$\pm$0.02 & 0.69$\pm$0.07
& 0.38$\pm$0.03 & 1.26$\pm$0.17 \\
20 & 1.07$\pm$0.07 & --- & 0.63$\pm$0.05 & 1.02$\pm$0.13
& 1.21$\pm$0.13 & 1.21$\pm$0.17 \\
21 & 1.06$\pm$0.05 & 1.04$\pm$0.27 & 0.31$\pm$0.02 & 0.73$\pm$0.08
& 0.54$\pm$0.04 & 1.43$\pm$0.18 \\
22 & 2.15$\pm$0.06 & --- & 1.09$\pm$0.06 & 0.53$\pm$0.07
& 0.21$\pm$0.02 & 2.59$\pm$0.66 \\
23 & 1.27$\pm$0.06 & --- & 0.63$\pm$0.04 & 1.76$\pm$0.17
& 0.80$\pm$0.07 & 1.59$\pm$0.29 \\
24 & 0.18$\pm$0.45 & --- & --- & 1.25$\pm$0.82
& 1.24$\pm$0.91 & 2.04$\pm$0.94 \\
25 & 1.20$\pm$0.06 & --- & 0.19$\pm$0.02 & 0.70$\pm$0.09
& 0.61$\pm$0.05 & 1.83$\pm$0.38 \\
26 & 0.75$\pm$0.05 & --- & 0.74$\pm$0.03 & 0.61$\pm$0.07
& 0.64$\pm$0.05 & 1.46$\pm$0.24 \\
27 & 1.08$\pm$0.04 & --- & 0.95$\pm$0.03 & 0.44$\pm$0.05
& 0.36$\pm$0.03 & 1.47$\pm$0.24 \\
28 & 1.19$\pm$0.19 & --- & 1.83$\pm$0.27 & 0.96$\pm$0.29
& 1.28$\pm$0.37 & 1.85$\pm$0.56 \\
29 & 0.85$\pm$0.15 & --- & --- & 0.93$\pm$0.22
& 0.90$\pm$0.22 & 1.70$\pm$0.45 \\
30 & 1.11$\pm$0.04 & 0.92$\pm$0.11 & 2.02$\pm$0.05 & 0.52$\pm$0.05
& 0.51$\pm$0.03 & 1.46$\pm$0.18 \\
31 & 0.59$\pm$0.08 & --- & 1.24$\pm$0.08 & 0.76$\pm$0.11
& 0.90$\pm$0.11 & 1.36$\pm$0.19 \\
32 & 1.25$\pm$0.08 & --- & 1.56$\pm$0.11 & 0.82$\pm$0.12
& 0.93$\pm$0.11 & 1.49$\pm$0.22 \\
33 & 0.78$\pm$0.19 & --- & 0.36$\pm$0.08 & 1.57$\pm$0.44
& 0.90$\pm$0.28 & 2.19$\pm$0.95 \\
34 & 1.04$\pm$0.06 & --- & 0.19$\pm$0.03 & 1.09$\pm$0.12
& 0.45$\pm$0.05 & 1.55$\pm$0.35 \\
35 & 1.57$\pm$0.17 & --- & --- & 1.80$\pm$0.45
& 0.74$\pm$0.21 & 3.10$\pm$1.98 \\
36 & 0.65$\pm$0.07 & 1.72$\pm$0.43 & 0.08$\pm$0.02 & 0.67$\pm$0.10
& 0.61$\pm$0.06 & 1.44$\pm$0.28 \\
37 & 0.77$\pm$0.29 & --- & --- & 0.70$\pm$0.33
& 0.49$\pm$0.25 & 1.87$\pm$1.58 \\
38 & 0.25$\pm$0.08 & --- & 0.26$\pm$0.05 & 0.92$\pm$0.13
& 0.38$\pm$0.06 & 0.97$\pm$0.29 \\
39 & 1.25$\pm$0.05 & --- & 0.15$\pm$0.01 & 0.88$\pm$0.08
& 0.71$\pm$0.04 & 1.46$\pm$0.19 \\[1mm]
\hline
\end{tabular}
\end{center}
\end{table*}

\subsection{Comparison of Results with Previous Studies}

\begin{figure}
\vspace{6.0mm}
\centerline{\includegraphics[width=7.0cm]{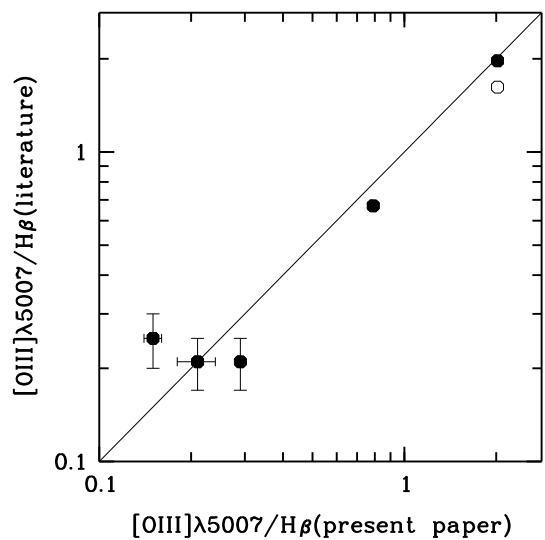}}
\caption{A comparison of relative fluxes in the line of oxygen 
[O\,{\sc iii}]$\lambda$5007\AA, obtained in the present work with the data 
from \citet{mccall1985} (black circles) and \citet{ferguson1998} (open 
circles) for common objects. The measurement errors, not exceeding the sizes 
of the signs in the figure are not shown.
\label{figure:compare}}
\end{figure}

We compared our estimates of relative intensity of the H~II region 
emission lines in the galaxy with the results from 
\citet{mccall1985,ferguson1998}. Four objects from our sample (one of 
which was observed twice) coincide with the H~II regions from the list 
of \citet{mccall1985}, one of these objects was also studied in 
\citet{ferguson1998} (Table~\ref{table:poshii}). For the reasons 
stated in the introduction, we did not compare our results with the 
spectrophotometric data \citep{belley1992}. Fig.~\ref{figure:compare} 
in a way of example gives a comparison of relative fluxes in the 
[O\,{\sc iii}]$\lambda$5007\AA \, line for the coinciding objects that 
were obtained by us and in \citet{mccall1985,ferguson1998}. The figure 
demonstrates a satisfactory agreement between them.

The differences between our results and those obtained by 
\citet{mccall1985,ferguson1998}, which are for some objects higher than 
the measurement errors, may be caused by the fact that the angular 
scales of the H~II regions in the nearby NGC~6946 galaxy are as a rule 
greater than the width of the slit, set for the spectroscopic 
observations. Different authors obtain the spectra of various parts of 
the H~II region with a slightly different chemical composition. Also note 
the problem, associated with the identification of H~II regions in the 
given stellar system: NGC~6946 is an example of a galaxy with a virtually 
complete absence of large stellar complexes. Its spiral arms are the 
chains of closely spaced H~II regions. A slightly different seeing during 
the observations may result in the differences in the deﬁnition of what 
actually constitutes an individual H~II region.

\subsection{Metallicity Gradient, Electron Temperature}

The metallicity variations in the H~II regions, depending on their 
galactocentric distance reflect the chemical evolution of the disk galaxies. 
To investigate the metallicity gradient in the galactic disk we need a 
sample of H~II regions, relatively evenly distributed over the galactocentric 
distances. Such samples of H~II regions with measured chemical compositions 
are available only for a limited number (about 50) of nearby galaxies 
\citep[see the following surveys:][]{garnett2002,pilyugin2004,moustakas2010}. 
We used the sample of 30 objects investigated in this study to determine the 
metallicity gradient in the disk of NGC~6946.

\begin{table}
\begin{center}
\caption{The oxygen, nitrogen abundances and electron temperature in 
the H~II regions.}
\label{table:abun}
\begin{tabular}{|c|c|c|c|c|}
\hline\hline
H~II    & $r/R_{25}$ & 12+         & 12+         & $t_{\rm NS}$,  \\
regions &            & $\log$(O/H) & $\log$(N/H) & $10^4$ K  \\
\hline
 1 & 0.40 & 8.73$\pm$0.04 & 7.77$\pm$0.05 & 0.64$\pm$0.01 \\
 2 & 0.49 & 8.46$\pm$0.01 & 7.74$\pm$0.02 & 0.81$\pm$0.01 \\
 6 & 0.22 & 8.58$\pm$0.02 & 7.87$\pm$0.03 & 0.72$\pm$0.01 \\
 8a & 0.17 & 8.62$\pm$0.02 & 7.83$\pm$0.03 & 0.69$\pm$0.01 \\
 8b & 0.17 & 8.59$\pm$0.01 & 7.79$\pm$0.03 & 0.72$\pm$0.01 \\
 9 & 0.17 & 8.67$\pm$0.03 & 7.73$\pm$0.05 & 0.68$\pm$0.01 \\
10 & 0.21 & 8.74$\pm$0.05 & 8.43$\pm$0.02 & 0.58$\pm$0.02 \\
11 & 0.25 & 8.68$\pm$0.03 & 8.40$\pm$0.01 & 0.61$\pm$0.01 \\
13 & 0.23 & 8.54$\pm$0.02 & 8.09$\pm$0.03 & 0.71$\pm$0.01 \\
14 & 0.29 & 8.67$\pm$0.02 & 7.97$\pm$0.02 & 0.66$\pm$0.01 \\
15a & 0.30 & 8.54$\pm$0.02 & 7.70$\pm$0.03 & 0.75$\pm$0.01 \\
15b & 0.30 & 8.68$\pm$0.02 & 7.98$\pm$0.04 & 0.67$\pm$0.01 \\
16 & 0.33 & 8.64$\pm$0.02 & 8.05$\pm$0.03 & 0.67$\pm$0.01 \\
17 & 0.32 & 8.63$\pm$0.02 & 8.11$\pm$0.03 & 0.67$\pm$0.01 \\
18 & 0.31 & 8.60$\pm$0.02 & 7.55$\pm$0.04 & 0.72$\pm$0.01 \\
19 & 0.38 & 8.60$\pm$0.01 & 7.80$\pm$0.03 & 0.73$\pm$0.01 \\
20 & 0.36 & 8.52$\pm$0.02 & 7.36$\pm$0.05 & 0.79$\pm$0.01 \\
21 & 0.39 & 8.61$\pm$0.01 & 7.78$\pm$0.02 & 0.71$\pm$0.01 \\
22 & 0.48 & 8.56$\pm$0.02 & 7.66$\pm$0.05 & 0.78$\pm$0.01 \\
23 & 0.48 & 8.46$\pm$0.01 & 7.89$\pm$0.03 & 0.79$\pm$0.01 \\
25 & 0.25 & 8.64$\pm$0.02 & 7.80$\pm$0.03 & 0.68$\pm$0.01 \\
26 & 0.37 & 8.52$\pm$0.01 & 7.50$\pm$0.03 & 0.80$\pm$0.01 \\
27 & 0.36 & 8.57$\pm$0.01 & 7.53$\pm$0.03 & 0.78$\pm$0.01 \\
28 & 0.38 & 8.53$\pm$0.05 & 7.53$\pm$0.12 & 0.87$\pm$0.02 \\
30 & 0.45 & 8.47$\pm$0.02 & 7.32$\pm$0.04 & 0.88$\pm$0.01 \\
31 & 0.49 & 8.49$\pm$0.02 & 7.32$\pm$0.04 & 0.84$\pm$0.01 \\
32 & 0.48 & 8.52$\pm$0.02 & 7.36$\pm$0.05 & 0.85$\pm$0.01 \\
33 & 0.19 & 8.50$\pm$0.05 & 7.89$\pm$0.09 & 0.75$\pm$0.02 \\
34 & 0.16 & 8.66$\pm$0.02 & 8.10$\pm$0.02 & 0.66$\pm$0.01 \\
36 & 0.23 & 8.73$\pm$0.03 & 7.93$\pm$0.02 & 0.63$\pm$0.01 \\
38 & 0.40 & 8.66$\pm$0.03 & 8.05$\pm$0.04 & 0.67$\pm$0.01 \\
39 & 0.32 & 8.64$\pm$0.01 & 7.88$\pm$0.02 & 0.67$\pm$0.01 \\[1mm]
\hline
\end{tabular}
\end{center}
\end{table}

The study of the radial distribution of nitrogen in galaxies is generally
given less attention than the research of the O/H gradient. However, the 
knowledge of the radial N/H gradient is important for the study of the 
chemical evolution of galaxies. Starting from the values of 
$12+\log({\rm O/H}) \ge 8.3$ the secondary nitrogen begins to 
dominate in the gas disk. Its abundance increases more rapidly than the 
oxygen content \citep{henry2000}. As a result, the variations of the 
nitrogen abundance with distance from the center of the galaxy has a 
greater amplitude than the variation in the oxygen content, and can be 
determined with a good accuracy, despite the fact that the fluxes in the 
nitrogen lines are usually measured with larger errors than those in the 
lines of oxygen. In addition, a comparative analysis of the O/H and N/H 
gradients in the galaxy can provide information on the delay of appearance 
of nitrogen in the interstellar medium with respect to oxygen 
\citep{maeder1992,vandenhoek1997,pagel1997book,pilyuginthuan2011}. 
Therefore, we determine here the radial gradients of both oxygen and 
nitrogen.

Radial distributions of metallicity are generally described by the 
following expressions: 
\begin{equation}
12+\log({\rm O/H}) = 12+\log({\rm O/H})_0+C_{\rm O/H}\times(r/R_{25}),
\label{equation:grado}
\end{equation}
where $12+\log({\rm O/H})_0$ is the relative abundance of oxygen, 
extrapolated to the center, $C_{\rm O/H}$ is the value of gradient of the 
radial decrease of relative oxygen abundance in the units of 
dex/$R_{\rm 25}$, $r$/$R_{\rm 25}$ is the galactocentric distance in the 
units of the galactic radius, measured from the isophote 25$^m$/arcsec$^2$ 
\citep{zaritsky1994,vanzee1998,pilyugin2004}. The radial distribution of 
relative nitrogen content is described similarly: 
\begin{equation}
12+\log({\rm N/H}) = 12+\log({\rm N/H})_0+C_{\rm N/H}\times(r/R_{25}).
\label{equation:gradn}
\end{equation}

\begin{figure}
\vspace{6.0mm}
\centerline{\includegraphics[width=7.0cm]{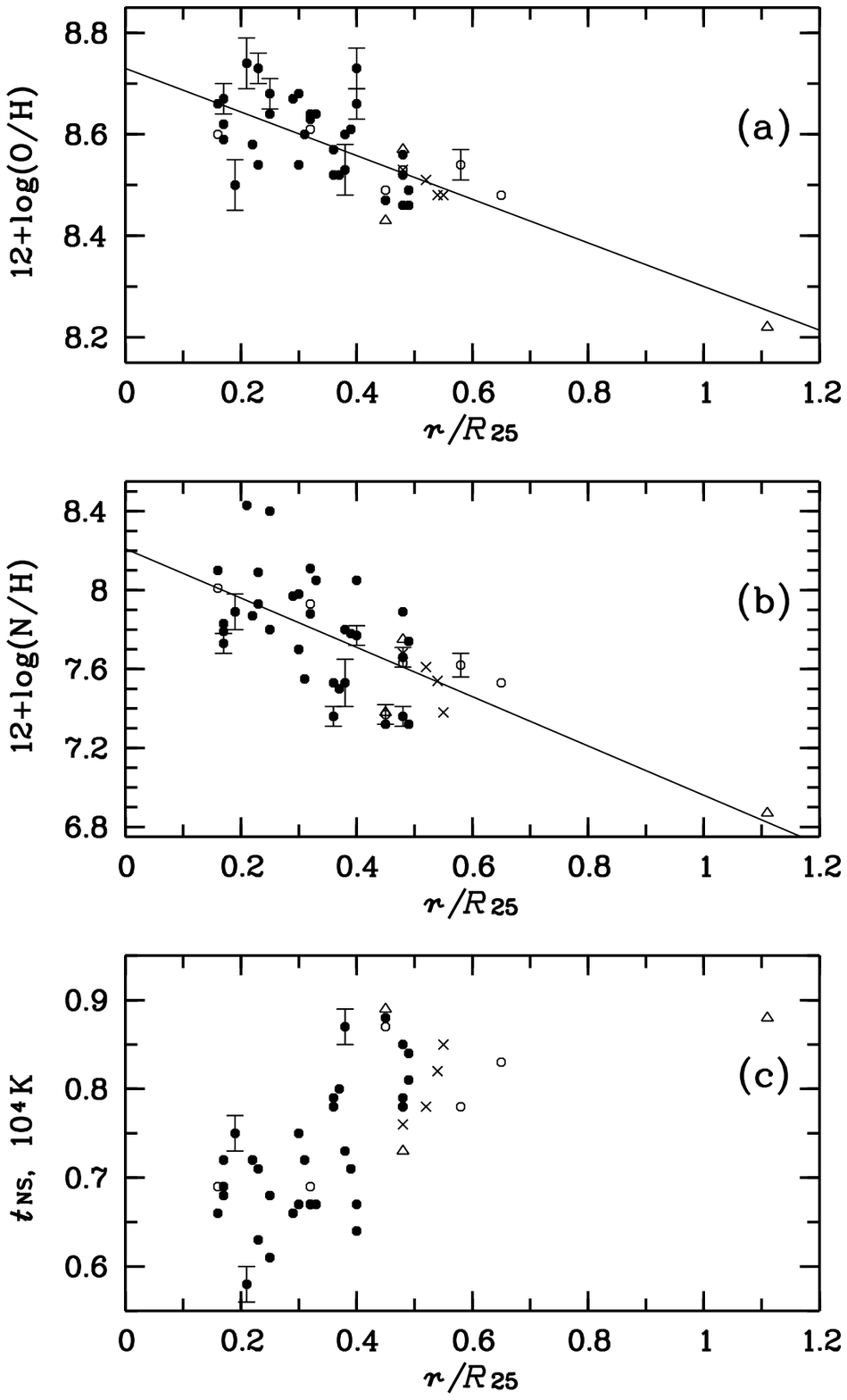}}
\caption{The diagrams of radial distribution of the relative abundance 
of oxygen (a), nitrogen (b) and the values of electron temperature 
$t_{\rm NS}$ (c) in the disk of NGC~6946. The open circles denote the 
objects, measured in \citet{mccall1985}, the open triangles are the 
measurements from \citet{ferguson1998}, the oblique crosses --- 
\citet{garcia2010}, while the black circles are the objects studied in this 
paper. The values determined with an accuracy of less than 
$\pm0.02$~dex (a), $\pm0.04$~dex (b) and $\pm100$~K (c) are accompanied 
with the error bars. The straight lines describe the radial gradient of the 
oxygen and nitrogen abundances.
\label{figure:grad}}
\end{figure}

Figs.~\ref{figure:grad}a and ~\ref{figure:grad}b show the radial 
distribution of relative oxygen and nitrogen abundances in the disk of 
NGC~6946, respectively. We can see that the actual spread of metallicity 
at a fixed radius exceeds the chemical composition measurement errors. 
The objects from \citet{mccall1985,ferguson1998} are also subject to this 
spread: one of the objects from the samples of 
\citet{mccall1985,ferguson1998} is identified with the object no.~30 from 
our sample and has the same chemical composition.

\begin{figure}
\vspace{7.0mm}
\centerline{\includegraphics[width=6.0cm]{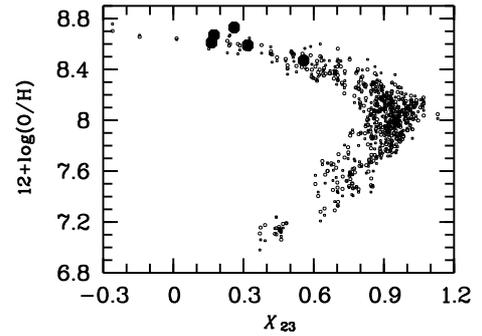}}
\caption{The $X_{23}$--O/H diagram for the sample of H~II regions according 
to \citet{pilyugin2010} (dots), \citet{pilyugin2012} (gray dots), and five 
objects investigated in the present study (black circles).
\label{figure:x23}}
\end{figure}

The numerical values of the sought coefficients in the expressions
(\ref{equation:grado}) and (\ref{equation:gradn}) were determined via the
least squares method for all the objects shown in Fig.~\ref{figure:grad}.
We obtained the following values of the
coefficients: \\
$12+\log({\rm O/H})_0 = 8.73\pm0.02$, \\
$C_{\rm O/H} = -0.43\pm0.06$ dex/$R_{\rm 25}$, \\
$12+\log({\rm N/H})_0 = 8.21\pm0.08$, \\
$C_{\rm N/H} = -1.25\pm0.19$ dex/$R_{\rm 25}$.

Fig.~\ref{figure:grad} shows that the radial distributions of the oxygen 
and nitrogen abundances are described by a single regression model 
throughout the entire disk of NGC~6946. The distribution of the nitrogen 
abundance varies faster relative to the radial distribution of oxygen.

The galaxy reveals an increase in the electron temperature of the H~II 
regions with distance from the galactic center (Fig.~\ref{figure:grad}c). 
A similar dependence is also observed for the H~II regions in the Galaxy 
\citep{churchwell1978,paladini2004,quireza2006,balser2011}. The 
temperature–-metallicity dependence for the H~II regions in NGC~6946 is 
rather well explained by the theoretical models of \citet{rubin1985}.

\section{DISCUSSION}

Figs.~\ref{figure:grad}a and ~\ref{figure:grad}b show a large dispersion 
of object positions relative to the regression model of the radial gradient 
of chemical composition. This is consistent with the results of previous 
studies \citep{mccall1985,ferguson1998,garcia2010} for the common and 
nearby objects. \citet{mccall1985,ferguson1998} have investigated the 
abundance of chemical elements in the MRS~4 (+182, +103) = FGW~6946A 
complex (corresponds to the complex 30 from our list), and obtained the 
metallicities, which coincide with our values within the errors 
(Fig.~\ref{figure:grad}). On the northern edge of the spiral arm (the 
eastern part of the galaxy) a giant stellar complex is located, it was 
investigated in \citet{garcia2010} and denominated as Knot~A. The 
estimations of the oxygen and nitrogen abundances in the complex, obtained 
in \citet{garcia2010}, proved to be similar 
($12+\log{\rm (O/H)}=8.48\pm0.01$, $12+\log{\rm (N/H)}=7.38\pm0.01$) with 
the O/H and N/H estimates, obtained in the present work for the most 
nearby to the Knot~A regions number 31 and 32 
(Figs.~\ref{figure:map}, \ref{figure:grad}, Table~\ref{table:abun}).

\begin{figure}
\vspace{7.0mm}
\centerline{\includegraphics[width=6.0cm]{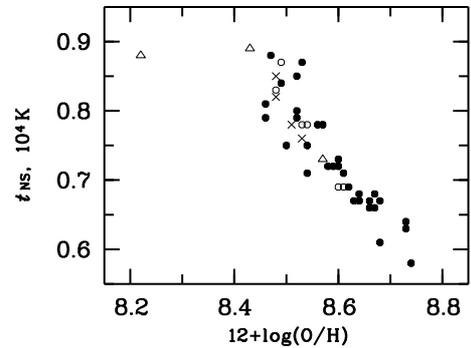}}
\caption{The dependence of the electron temperature on the relative 
abundance of oxygen in the H~II regions. The designations are the same 
as in Fig.~\ref{figure:grad}.
\label{figure:toh}}
\end{figure}

A large dispersion is most likely due to the conditions of observations of 
the H~II regions in this galaxy. NGC~6946 is a sufficiently close galaxy, 
and the nebulae in it have relatively large angular dimensions. In the 
case if the spectrograph slit has only captured the central part of the 
nebula, where the nitrogen is doubly ionized, the NS-calibration used here 
leads to an underestimation of the oxygen content. If the spectrograph 
slit captures the peripheral part of the nebula, where the doubly ionized 
nitrogen is absent, the NS-calibration yields an exaggerated oxygen 
abundance. In order to test the reliability of the obtained oxygen 
abundance estimates, we checked the positions of the investigated objects 
on other diagrams.

Based on the usual assumption that in the H~II regions having similar 
intensities of strong emission lines there should be approximately the same 
physical conditions and chemical composition, we have compared in 
Fig.~\ref{figure:x23} the positions of objects, studied in this research 
(the black circles) with a sample of calibration areas from 
\citet{pilyugin2010} (the black dots). The gray dots show the H~II regions 
for which the oxygen abundance was determined via the recently proposed 
C-method \citep{pilyugin2012}. We determined the parameter 
$X_{23} \equiv \log R_{23} \equiv \log$~(([O\,{\sc ii}]$\lambda$3727+
$\lambda$3729\AA+[O\,{\sc iii}]$\lambda$4959\AA+[O\,{\sc iii}]$\lambda$5007\AA)/H$\beta$) 
for five H~II regions in NGC~6946, where we were able to measure the 
[O\,{\sc ii}]$\lambda$3727+$\lambda$3729\AA \, line (Table~\ref{table:flux1}) 
as $\log$~(([O\,{\sc ii}]$\lambda$3727+$\lambda$3729\AA+1.33[O\,{\sc iii}]$\lambda$5007\AA)/H$\beta$).

\begin{figure}
\vspace{7.0mm}
\centerline{\includegraphics[width=6.0cm]{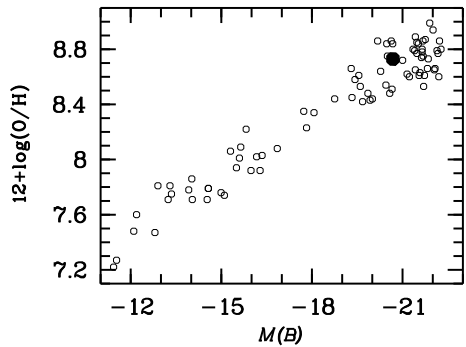}}
\caption{The luminosity–-central metallicity diagram. The open circles 
represent the data from \citet{pilyugin2007} \citep[with the addition of the 
results from][]{gusev2012}, the black circle marks the position of 
NGC~6946 according to this paper.
\label{figure:mb_oh}}
\end{figure}

Fig.~\ref{figure:toh} examines the relationship between the electron 
temperature and oxygen abundance in the studied regions, indicating that 
the electron temperature in the nebula essentially depends on the cooling 
of gas via irradiation in the oxygen lines. Here, like in 
Fig.~\ref{figure:grad}c, a significantly underestimated electron temperature 
of the object from the sample of \citet{ferguson1998} can be seen. Since 
this object was located far at the edge of the galactic disk 
($R/R_{25} > 1$), the account of this object can effect the choice of the 
radial the gradient of chemical composition of the galaxy. Without this 
object the gradient becomes a little flatter, 
$-0.39\pm0.08$ dex/$R_{\rm 25}$ in the case of radial distribution of 
oxygen, and a little steeper $-1.30\pm0.22$ dex/$R_{\rm 25}$ in the case of 
radial distribution of nitrogen.

The luminosity -– central metallicity diagram was built by 
\citet{pilyugin2007}. The oxygen and nitrogen abundances in 
\citet{pilyugin2007} and in the present work were determined using the 
methods that are consistent with the metallicity scale, based on the $T_{e}$ 
method, hence they can be compared. Fig.~\ref{figure:mb_oh} demonstrates the 
position of the central metallicity estimates of NGC~6946 obtained here 
compared with other galaxies. It is clear from the graph that the central 
metallicity obtained in this work is typical for the galaxy of a given 
luminosity.

\section{CONCLUSIONS}

Using the SCORPIO focal reducer in the multislit mode we have performed 
the spectroscopy of 39 H~II regions of the NGC~6946 galaxy. The estimates 
of absorption were obtained for these regions.

Having applied the ''strong line'' method the electron temperatures, oxygen 
and nitrogen abundances were determined for 30 H~II regions. The radial 
gradients of O/H, N/H, and the electron temperature were constructed.

The radial decrease of the oxygen and nitrogen abundances amounted to \\
$12+\log({\rm O/H}) = (8.73\pm0.02)-(0.43\pm0.06)r/R_{\rm 25}$ and \\
$12+\log({\rm N/H}) = (8.21\pm0.08)-(1.25\pm0.19)r/R_{\rm 25}$.

\begin{acknowledgements}
A.~S.~G. thanks A.~Yu.~Knyazev (South African Astronomical Observatory) for 
his advice on the spectral data reduction, V.~L.~Afanasiev and 
A.~V.~Moiseev (SAO RAS) for the assistance with the BTA observations and 
valuable advice, A.~V.~Zasov, B.~P.~Artamonov (SAI MSU) and L.~S.~Pilyugin 
(MAO NASU) for the productive discussion of the results. The paper used 
the data from the electronic HyperLeda database (http://leda.univ-lyon1.fr). 
The research was supported by the Russian Foundation for Basic Research 
(project nos. 08--02--01323, 10--02--91338, and 12--02--00827). The 
observations were carried out at the BTA telescope with the financial 
support of the Ministry of Education and Science of Russian Federation 
(state contracts no. 14.518.11.7070, 16.518.11.7073).
\end{acknowledgements}


\begin{thebibliography}{}

\bibitem[\protect\citeauthoryear{Afanasiev \& Moiseev}{2005}]{afanasiev2005}
Afanasiev L.V., Moiseev A.V., 2005, Astron. Lett., 31, 193

\bibitem[\protect\citeauthoryear{Baldwin et al.}{1981}]{baldwin1981}
Baldwin J.A., Phillips M.M., Terlevich R., 1981, PASP, 93, 5

\bibitem[\protect\citeauthoryear{Balser et al.}{2011}]{balser2011}
Balser D.S., Rood R.T., Bania T.M., Anderson L.D., 2011, ApJ, 738, 27

\bibitem[\protect\citeauthoryear{Belley \& Roy}{1992}]{belley1992}
Belley J., Roy J.-R., 1992, ApJS, 78, 61

\bibitem[\protect\citeauthoryear{Boomsma et al.}{2008}]{boomsma2008}
Boomsma R., Oosterloo T.A., Fraternali F., et al., 2008, A\&A, 490, 555

\bibitem[\protect\citeauthoryear{Bresolin et al.}{2005}]{bresolin2005}
Bresolin F., Schaerer D., Conz\'{a}lez Delgado R.M., Stasi\'{n}ska G., 
2005, A\&A, 441, 981

\bibitem[\protect\citeauthoryear{Bresolin et al.}{2009}]{bresolin2009}
Bresolin F., Gieren W., Kudritzki R.-P., Pietrzy\'{n}ski G., et al., 
2009, ApJ, 700, 309

\bibitem[\protect\citeauthoryear{Churchwell et al.}{1978}]{churchwell1978}
Churchwell E., Smith L.F., Mathis J., et al., 1978, A\&A, 70, 719

\bibitem[\protect\citeauthoryear{Dutil \& Roy}{1999}]{dutil1999}
Dutil D.R., Roy J.-R., 1999, ApJ, 516, 62

\bibitem[\protect\citeauthoryear{Efremov et al.}{2002}]{efremov2002}
Efremov Yu.N., Pustilnik S.A., Kniazev A.Y., et al., 2002, A\&A, 389, 855

\bibitem[\protect\citeauthoryear{Efremov et al.}{2007}]{efremov2007}
Efremov Yu.N., Afanasiev V.L., Alfaro E.J., et al., 2007, MNRAS, 382, 481

\bibitem[\protect\citeauthoryear{Efremov et al.}{2011}]{efremov2011}
Efremov Yu.N., Afanasiev V.L., Egorov O.V., 2011, Astroph. Bull., 66, 304

\bibitem[\protect\citeauthoryear{Ellison et al.}{2008}]{ellison2008}
Ellison S.L., Patton D.R., Simard L., McConnachie A.W., 
2008, AJ, 135, 1877

\bibitem[\protect\citeauthoryear{Elmegreen et al.}{2000}]{elmegreen2000}
Elmegreen B.G., Efremov Yu.N., Larsen S., 2000, ApJ, 535, 748

\bibitem[\protect\citeauthoryear{Ferguson et al.}{1998}]{ferguson1998}
Ferguson A.M.N., Gallagher J.S., Wyse R.F.G., 1998, AJ, 116, 673

\bibitem[\protect\citeauthoryear{Garc\'{i}a-Benito et al.}{2010}]{garcia2010}
Garc\'{i}a-Benito R., D\'{i}az A., Hagele G.F., et al., 2010, MNRAS, 
408, 2234

\bibitem[\protect\citeauthoryear{Garnett}{2002}]{garnett2002}
Garnett D.R., 2002, ApJ, 581, 1019

\bibitem[\protect\citeauthoryear{Gusev et al.}{2012}]{gusev2012} 
Gusev A.S., Pilyugin L.S., Sakhibov F., et al., 2012, MNRAS, 424, 1930

\bibitem[\protect\citeauthoryear{Guti\'{e}rrez \& Beckman}{2010}]{gutierrez2010}
Guti\'{e}rrez L., Beckman J.E., 2010, ApJ, 710, L44

\bibitem[\protect\citeauthoryear{Henry et al.}{2000}]{henry2000}
Henry R.B.C., Edmunds M.G., K\'{o}ppen J., 2000, ApJ, 541, 660

\bibitem[\protect\citeauthoryear{Hodge}{1967}]{hodge1967}
Hodge P.W., 1967, PASP, 79, 297

\bibitem[\protect\citeauthoryear{Izotov et al.}{1994}]{izotov1994}
Izotov Y.I., Thuan T.X., Lipovetsky V.A., 1994, ApJ, 435, 647

\bibitem[\protect\citeauthoryear{Karachentsev et al.}{2000}]{karachentsev2000}
Karachentsev I.D., Shavrina M.E., Huchtmeier W.K., 2000, A\&A, 362, 544
 
\bibitem[\protect\citeauthoryear{Kartasheva \& Chounakova}{1978}]{kartasheva1978}
Kartasheva T.A., Chunakova N.M., 1978, Izv. SAO, 10, 44

\bibitem[\protect\citeauthoryear{Kauffmann et al.}{2003}]{kauffmann2003}
Kauffmann G., Heckman T.M., Tremonti C., et al., 2003, MNRAS, 346, 1055

\bibitem[\protect\citeauthoryear{Kennicutt}{1984}]{kennicutt1984}
Kennicutt R.C., 1984, ApJ, 287, 116

\bibitem[\protect\citeauthoryear{Kennicutt et al.}{2003}]{kennicutt2003}
Kennicutt R.C., Bresolin F., Garnett D.R., 2003, ApJ, 591, 801

\bibitem[\protect\citeauthoryear{Kewley et al.}{2001}]{kewley2001}
Kewley L.J., Dopita M.A., Sutherland R.S., et al., 2001, ApJ, 556, 121

\bibitem[\protect\citeauthoryear{L\'{o}pez-S\'{a}nchez \& Esteban}{2010}]{lopezsanchez2010}
L\'opez-S\'anchez \'A.R., Esteban C., 2010, A\&A, 517, A85

\bibitem[\protect\citeauthoryear{Maeder}{1992}]{maeder1992}
Maeder A., 1992, A\&A, 264, 105

\bibitem[\protect\citeauthoryear{McCall et al.}{1985}]{mccall1985}
McCall M.L., Rybski P.M., Shields G.A., 1985, ApJS, 57, 1

\bibitem[\protect\citeauthoryear{Moustakas et al.}{2010}]{moustakas2010}
Moustakas J., Kennicutt R.C. (Jr), Tremonti C.A., et al., 2010, ApJS, 
190, 233

\bibitem[\protect\citeauthoryear{Oke}{1990}]{oke1990}
Oke J.B., 1990, AJ, 99, 1621

\bibitem[\protect\citeauthoryear{Osterbrock}{1989}]{osterbrock1989}
Osterbrock D.E., 1989, Astrophysics of gaseous nebulae and active galactic 
nuclei (Mill Valley, CA: University Science Books), 422~p.

\bibitem[\protect\citeauthoryear{Pagel}{1997}]{pagel1997book}
Pagel B.E.J., 1997, Nucleosynthesis and Chemical Evolution of Galaxies 
(Cambridge: Cambridge Univ. Press), 392~p.

\bibitem[\protect\citeauthoryear{Paladini et al.}{2004}]{paladini2004}
Paladini R., Davies R.D., DeZotti G., 2004, MNRAS, 347, 237

\bibitem[\protect\citeauthoryear{Paturel et al.}{2003}]{paturel2003}
Paturel G., Petit C., Prugniel Ph., et al., 2003, A\&A, 412, 45

\bibitem[\protect\citeauthoryear{Pilyugin}{2003}]{pilyugin2003}
Pilyugin L.S., 2003, A\&A, 399, 1003

\bibitem[\protect\citeauthoryear{Pilyugin \& Mattsson}{2011}]{pilyuginmattsson2011}
Pilyugin L.S., Mattsson L., 2011, MNRAS, 412, 1145

\bibitem[\protect\citeauthoryear{Pilyugin \& Thuan}{2011}]{pilyuginthuan2011}
Pilyugin L.S., Thuan T.X., 2011, ApJ, 726, L23

\bibitem[\protect\citeauthoryear{Pilyugin et al.}{2004}]{pilyugin2004}
Pilyugin L.S., V\'{\i}lchez J.M., Contini T., 2004, A\&A, 425, 849

\bibitem[\protect\citeauthoryear{Pilyugin et al.}{2007}]{pilyugin2007}
Pilyugin L.S., Thuan T.X., V\'{\i}lchez J.M., 2007, MNRAS, 376, 353

\bibitem[\protect\citeauthoryear{Pilyugin et al.}{2010}]{pilyugin2010}
Pilyugin L.S., V\'{i}lchez J.M., Thuan T.X., 2010, ApJ, 720, 1738

\bibitem[\protect\citeauthoryear{Pilyugin et al.}{2012}]{pilyugin2012}
Pilyugin L.S., Grebel E.K., Mattsson L., 2012, MNRAS, 424, 2316

\bibitem[\protect\citeauthoryear{Quireza et al.}{2006}]{quireza2006}
Quireza C., Rood R.T., Bania T.M., et al., 2006, ApJ, 653, 1226

\bibitem[\protect\citeauthoryear{Roy et al.}{1996}]{roy1996}
Roy J.-R., Belley J., Dutil Y., Martin P., 1996, ApJ, 460, 294

\bibitem[\protect\citeauthoryear{Rubin}{1985}]{rubin1985}
Rubin R.H., 1985, ApJS, 57, 349

\bibitem[\protect\citeauthoryear{Sakhibov}{2004}]{sakhibov2004}
Sakhibov F.H., 2004, Doctor's Dissertation in Physics and Mathematics 
(Moscow: MSU)

\bibitem[\protect\citeauthoryear{Storey \& Zeippen}{2000}]{storey2000}
Storey P.J., Zeippen C.J., 2000, MNRAS, 312, 813

\bibitem[\protect\citeauthoryear{van den Hoek \& Groenewegen}{1997}]{vandenhoek1997}
van den Hoek L.B., Groenewegen M.A.T., 1997, A\&AS, 123, 305

\bibitem[\protect\citeauthoryear{van Zee et al.}{1998}]{vanzee1998}
van Zee L., Salzer J.J., Haynes M.P., et al., 1998, AJ, 116, 2805

\bibitem[\protect\citeauthoryear{Zaritsky et al.}{1994}]{zaritsky1994}
Zaritsky D., Kennicutt R.C., Huchra J.P., 1994, ApJ, 420, 87

\end{thebibliography}
\end{document}